\documentclass[12pt]{iopart}

\usepackage{graphicx}
\usepackage{url}
\usepackage{iopams} 
\newcommand{\BE}{\begin{equation}}
\newcommand{\EE}{\end{equation}}
\expandafter\let\csname equation*\endcsname\relax
\expandafter\let\csname endequation*\endcsname\relax
\usepackage{amsmath,amssymb,amscd,latexsym,amsthm,mathrsfs}
\def\key{(\ref{eq:key}) }
\def\V{{\mathcal V}}
\def\N{{\mathcal N}}
\def\e{{\rm e}}
\def\m{{\rm m}}
\usepackage{hyperref}
\usepackage{subcaption}
\usepackage{ tikz}
\begin{document}

\title {\Large New scaling laws for self-avoiding walks: bridges and worms }

\author{Bertrand Duplantier$^\dag$, Anthony J Guttmann$^\ddag$}
\address{$^\dag$ Institut de Physique Th\'eorique, Universit\'e Paris-Saclay, CEA, CNRS, B\^at. 774, Orme des Merisiers, CEA/Saclay
 F-91191 Gif-sur-Yvette, {\sc France}}
\address{$^\ddag$ School of Mathematics and Statistics,
The University of Melbourne, Victoria 3010, {\sc Australia}}

\setcounter{footnote}{0}

\begin{abstract}
We show how the theory of the critical behaviour of $d$-dimensional polymer networks gives a scaling relation for self-avoiding {\em bridges} that relates the critical exponent for bridges $\gamma_b$ to that of terminally-attached self-avoiding arches, $\gamma_{11},$ and the {correlation} length exponent $\nu.$ We find $\gamma_b = \gamma_{11}+\nu.$ In the case of the special transition, we find 
$\gamma_b({\rm sp}) = \frac{1}{2}[\gamma_{11}({\rm sp})+\gamma_{11}]+\nu.$ We provide compelling numerical evidence for this result in both two- and three-dimensions. Another subset of SAWs, called {\em worms}, are defined as the subset of SAWs whose origin and end-point have the same $x$-coordinate. We give a scaling relation for the corresponding critical exponent $\gamma_w,$ which is $\gamma_w=\gamma-\nu.$ This too is supported by enumerative results in the two-dimensional case. 

\end{abstract}

 \vspace{5mm}

Dedicated to the memory of Vladimir Rittenberg.\\

\section{Introduction}
\begin{figure}
\centering
\includegraphics[angle=+90,scale =0.32] {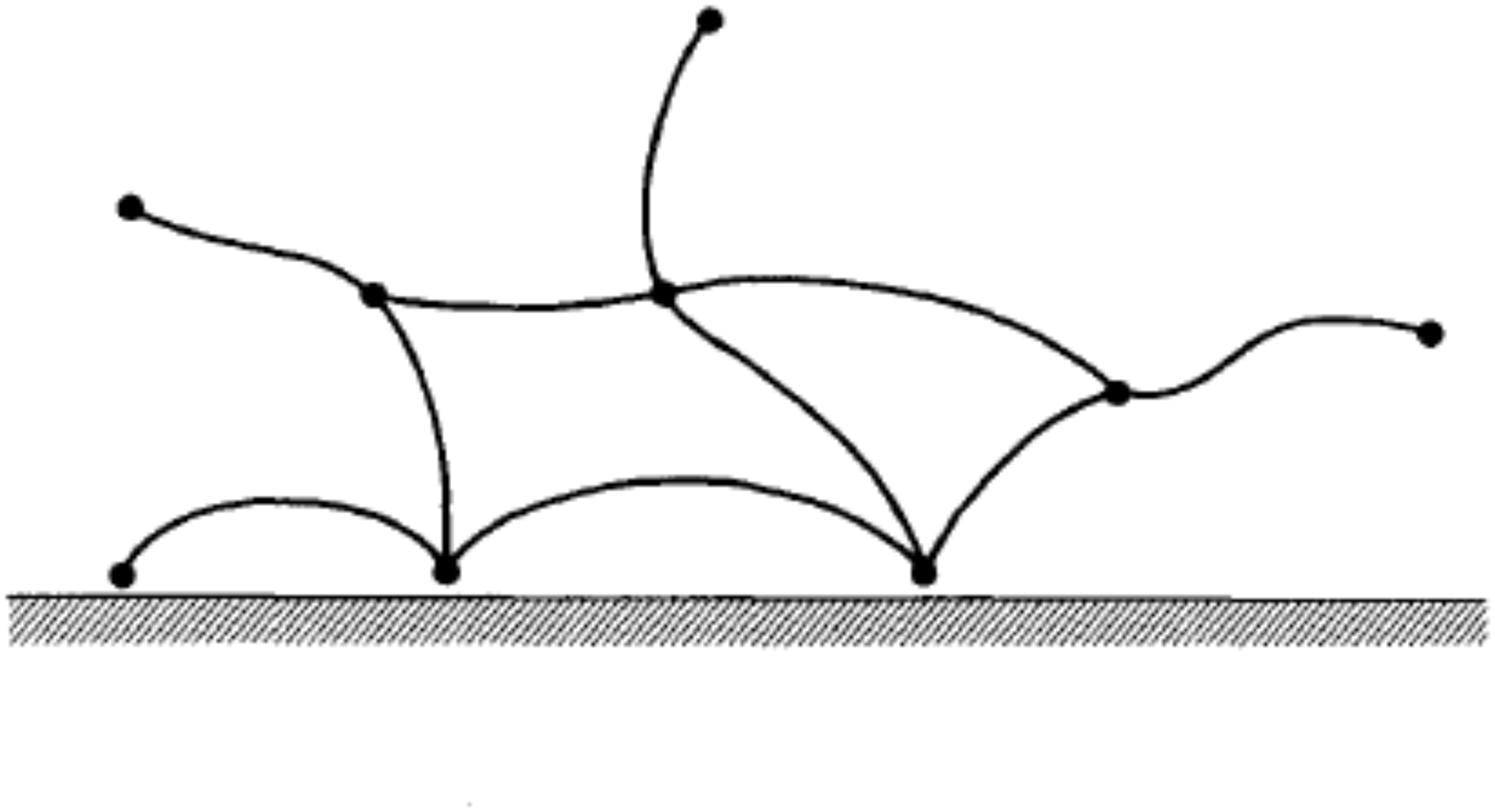}
 \caption{A polymer network ${\mathcal G}$ interacting with a surface. This network has 3 surface vertices, $n_1^S=1$ with one leg, $n_3^S=2$  with three legs; 6 bulk vertices, of which $n_1=3$ have one line, $n_3=2$ have three lines, and $n_4=1$ has four lines attached. To eliminate overall translational invariance, one (and only one) vertex is fixed along the surface. The network has $\mathcal N=10$  lines.}
 \label{fig:network}
\end{figure}

More than thirty years ago, an exhaustive treatment of the critical exponents of self-avoiding polymer networks in the bulk was proposed by one of us (BD) \cite{BD86}, and generalised to the boundary case in joint work with H Saleur \cite{DS86} (see also Ref. \cite{D89}). A typical such network is shown in Fig. \ref{fig:network}. 

A key result  
is that the configurational critical exponent associated with such a (monodisperse) $d$-dimensional polymer network ${\mathcal G}$ is given by
\begin{equation} \label{eq:key}
\gamma_{\mathcal G} = \nu \left[d{\mathcal V}+(d-1)({\mathcal V}_S-1) - \sum_{L \ge 1} (n_Lx_L+n_L^Sx_L^S)\right]-({\mathcal N}-1).
\EE
Here ${\mathcal V}$ is the number of bulk vertices, ${\mathcal V}_S$ is the number of surface vertices, $n_L$ is the number of $L$-leg vertices floating in the bulk, while $n_L^S$ is the number of $L$-leg vertices on or constrained close to, the surface. ${\mathcal N}$ denotes the number of chains in the network.  

The intuitive meaning of formula \eqref{eq:key} for such a configurational  exponent $\gamma_{\mathcal G}$ is clear: the first two terms correspond, {via the correlation lengh exponent $\nu$,} to the Euclidean phase space of the (bulk and surface) vertices of the network, the $x_L$ and $x_L^S$ exponents correspond to the reduction in phase space induced by both linkage and self- and mutual avoidance in each $L$-star vertex, and the last term, $\mathcal N-1$, corresponds to the constraint of \emph{monodispersity} in the $\mathcal N$ arms of the network ({\it i.e.}, their respective lengths, or monomer numbers,  all scale similarly).

For an unconstrained network in the bulk, with no surface, $d{\mathcal V}$ must be replaced by $d{\mathcal V}-1,$ as otherwise one is counting all translates. {A non-trivial result, of course, is that such a reduction to individual vertices holds true;  it can be obtained in 2 dimensions from \emph{conformal field theory} \cite{BD86,DS86,D89}, or from a \emph{two-dimensional quantum gravity} approach \cite{DK88,DK90}, and in generic dimension $d$, from \emph{renormalization theory} \cite{D89,SFLD92}.} 

In two-dimensions, the bulk conformal weights $(\frac{1}{2})x_L,$ and surface conformal weights $x_L^S,$  associated with $L-$vertices, are explicitly given by 
\begin{equation} \label{eq:xnorm}
x_L=\frac{1}{48}(3L-2)(3L+2), \,\,\,\,\,\,  x_L^S=\frac{1}{8}L(3L+2),
\end{equation}
with $1/\nu=2-x_2=4/3$. (See Refs \cite{Nienhuis82,Nienhuis87} for the $L=1,2$ bulk cases, Ref. \cite{Cardy84} for the $L=1$ boundary case, and \cite{S86,S87,DK88,BB89,D04} for the general bulk case, and \cite{DS86,BS93,D04} for the general boundary case.) In $d$-dimensions, they have the general form \cite{D89},
$$x_L=x^{\rm B}_L+x'_L,\,\,\,\,\,\, x^S_L=x^{S,{\rm B}}_L+x'^S_L,$$
where the first terms are the Brownian scaling dimensions, $x^{\rm B}_L:=\frac{L}{2}(d-2)$ and  $x^{S,{\rm B}}_L:=\frac{L}{2}d$, whereas the second ones, $x'_L,\, x'^S_L$, represent the anomalous contributions from self- and mutual avoidance. At first order in $\varepsilon:=4-d$, the latter are \cite{D89}
$$x'_L=\frac{\varepsilon}{8}L(L-1)+\left(\frac{\varepsilon}{8}\right)^2\frac{L}{4}\left(-8L^2+33L-23\right)+O(\varepsilon^3),$$ and $x'^S_L=\frac{\varepsilon}{8}L(L-2)+O(\varepsilon^2)$ \cite{D89}, while being also explicitly known to next order in $\varepsilon$ \cite{SFLD92,FH97}. For $d\geq 4$, $x'_L=0,x'^S_L=0$, with known logarithmic corrections to the Brownian network partition function for $d=4$ \cite{D89}. 

Note that Eq. \eqref{eq:key} naturally holds in the case of random walks or Brownian chains, for which $\nu=1/2$ and the scaling exponents take the above mentioned  Brownian values $x_L^{\rm B}, x_L^{S,{\rm B}}$. It also holds   in the mixed case of a network made of \emph{mutually-avoiding} (M) random walks or Brownian chains, for which the bulk and surface scaling exponents are in two-dimensions \cite{DKw88,D98,LSW01a,LSW01b} $$x_L^{\rm M}=\frac{1}{12}(4L^2-1),\,\,\, x_L^{S,\rm M}=\frac{1}{3}L(2L+1).$$
In $d=4-\varepsilon$ dimensions the bulk exponent is \cite{BDCMP87}
$$x_L^{\rm M}=x_L^{\rm B}+\frac{\varepsilon}{4}L(L-1)-\left(\frac{\varepsilon}{4}\right)^2L(L-1)(2L-5)+O(\varepsilon^3),$$
while for $d\geq 4$, $x_L^{\rm{M}}=x_L^{\rm B}$ and $x_L^{S,\rm M}=x_L^{S,\rm B}$, with known logarithmic corrections at $d=4$  \cite{BDCMP87}.

If there is an attractive surface fugacity $a=\exp(-\epsilon/k_BT),$ where $\epsilon$ is the energy associated with a monomer of the walk lying in the surface, then this has no effect on the critical point or critical exponent of the network provided that $a$ is less than its critical value $a_c > 1,$ where $a_c$ is called the {\em critical fugacity}. Precisely at the critical fugacity, the exponent changes discontinuously. This is called the {\em special transition}, and the corresponding expression to Eq. (\ref{eq:key}) follows straightforwardly by replacing the {scaling dimensions} $x_L^S$ \eqref{eq:xnorm} by their values at the special transition, obtained in Refs. \cite{FS94,BY95} (see \cite{GB89} for the $L=1$ case),
\begin{equation}\label{eq:xsurf}
x_L^S({\rm sp})=\frac{3}{8}(L+1)^2-\frac{3}{2}(L+1)+\frac{35}{24}.
\end{equation}
This extension to the case of polymer networks at the special transition in two-dimensions was given by Batchelor, Bennett-Wood and Owczarek \cite{BDO98}, who also studied the mixed case of some surface vertices being at the critical fugacity and others not. For $a > a_c$ the location of the critical point varies monotonically with $a,$ and the exponents change to integers, corresponding to poles in the generating function. In this paper we will only consider the situation $a \le a_c,$ which corresponds to the most interesting physics.

 We first show how the key result \key reproduces known exponent values and scaling laws for self-avoiding walks (SAWs) and their surface restricted counterparts, and then show how this can be extended to handle the case of bridges.
\subsection{Bulk and surface self-avoiding walks.}

A self-avoiding walk (SAW) on a lattice is an open, connected path on the lattice that does not revisit any vertex that has been previously visited. Walks are considered distinct if they are not translates of one another. If there are $c_n$ walks of length $n,$ each walk occurs with equal probability $1/c_n.$  It is known that $\lim_{n\to\infty} n^{-1}  \log c_n = \log \mu$ exists \cite{HM54}, where $\mu$ is the 
\emph{growth constant} of self-avoiding walks on the lattice.

While our primary result holds for all regular lattices, our numerical work will be confined to SAWs on the
 $d$-dimensional hypercubic lattice ${\mathbb Z}^d,$ with the vertices having 
integer coordinates $\{x_1^{(i)},x_2^{(i)}, \cdots, x_d^{(i)} \},$ for $i=0,1, \cdots,n.$ 

An $n$-step \emph{bridge} is a self-avoiding walk in the upper half-space through the origin that starts 
at the origin and is constrained (i) to have co-ordinate $x_d^{(i)} \ge 0$ for all 
$0 \le i \le n,$ and (ii) its end-point $\{ x_1^{(n)},x_2^{(n)}, \cdots, x_d^{(n)} \}$ is the unique point with maximal coordinate $x_d^{(i)}.$ That is to say, $x_d^{(n)}>x_d^{(i)}$ for all $0 \le i < n.$ The number of $n$-step bridges from 
the origin is denoted by $b_n$.  It is known that 
$\lim_{n\to\infty} n^{-1} \log b_n = \log \mu,$ where $\mu$ is unchanged from the corresponding value for SAWs \cite{HW62}. The generating function for bridges is $B(x)=\sum _{n \ge 0} b_n x^n$, and we denote bridges spanning a strip of width $T$ as $B_T(x)=\sum _{n \ge 0} b_n(T) x^n.$ One of the few rigorous results known about bridges, proved by Beaton et al. in \cite{BBDDG14}, is that $$\lim_{T \to \infty} B_T(1/\mu) = 0.$$

A terminally attached walk (TAW) is a SAW with one end anchored in the surface, but with the rest of the walk free in the upper half-space. Clearly TAWs are a superset of bridges, and a subset of SAWs, so have the same growth constant.

The next subset of SAWs we wish to consider are {\em arches}, which are SAWs in the upper half-plane with both the origin and end-point constrained to lie in the $(d-1)$-dimensional surface. That is to say, $x_d^{(0)} = 0 = x_d^{(n)}.$ As the number of arches is bounded above by the number of SAWs and below by the number of self-avoiding polygons (SAPs), which are known to have the same growth constant as SAWs \cite{H61}, it follows that arches also have the same growth constant as SAWs.

The last subset we consider is that of {\em worms}.  A worm is a SAW with origin and end-point coordinates satisfying $x_d^{(0)} = 0 = x_d^{(n)}.$ This is also a condition imposed on arches, but without the upper half-plane constraint satisfied by arches. These are clearly a subset of SAWs and a superset of arches, so again have the same growth constant  as SAWs. The number of $n$-step worms is denoted $w_n.$

The results on growth constants are essentially the only results that have been proved. Nevertheless, it is universally accepted that the asymptotic behaviour of the above objects is given by:
$$c_n \sim A \cdot \mu^n \cdot n^{\gamma-1},$$
$$b_n \sim B \cdot \mu^n \cdot n^{\gamma_b-1},$$
$$t_n \sim   C \cdot\mu^n \cdot n^{\gamma_1-1},$$
$$a_n \sim  D \cdot\mu^n \cdot n^{\gamma_{11}-1},$$
$$w_n \sim  E \cdot\mu^n \cdot n^{\gamma_{w}-1},$$
for SAWs, bridges, TAWs, arches and worms respectively.

In two dimensions, it is believed that $\gamma = 43/32$ \cite{Nienhuis82},  that $\gamma_1 = 61/64$  and that $\gamma_{11} = -3/16$ \cite{Cardy84}. As far as we are aware, until very recently there have been no published estimates for $\gamma_b$ or $\gamma_w,$ but as pointed out in \cite{CCG16}, last century one of us (AJG) estimated the value of this exponent in the two-dimensional case by series analysis to be $9/16.$ Subsequently, much longer series were calculated by Iwan Jensen, which enabled this estimate to be conjectured with much greater confidence. Somewhat later, Alberts and Madras (private communication) obtained the estimate $\gamma_b=9/16$ for two-dimensional bridges using SLE arguments, subject to certain unproven assumptions, but this work was never published.

In \cite{DGK11} both SAWs spanning a strip and bridges were discussed, and comparisons made with conjectured results from ${\rm SLE}_{8/3}.$ By arguing that the {probability  measure of bridges starting at 0 and ending
at  $x + iy$ should be given (up to normalisation) by the explicit function,  $\left[y\cosh ({\pi x}/{2y})\right]^{-5/4}$ \cite{DGK11}, 
Lawler (private communication) has provided a simple heuristic argument that predicts $\gamma_b = (3/4)^2 = 9/16,$ as above.

Another critical exponent that needs to be defined is that characterising the length of a SAW. Any standard measure of length, such as mean-square end-to-end distance, mean-square radius of gyration, squared caliper span etc., all behave as $const. \times n^{2\nu},$ where in two dimensions it is accepted that $\nu = 3/4$  \cite{Nienhuis82}.

These exponents are not all independent. There is a scaling relation, due to Barber \cite{B73},
\begin{equation}\label{barb}
2\gamma_1 - \gamma_{11} = \gamma + \nu,
\end{equation}
 which holds independent of dimension, and clearly links the exponents. As we show below, this also follows from \key.

For TAWs and arches (as well as for bridges and worms), $L=1,$ and from (\ref{eq:xnorm}) and (\ref{eq:xsurf})  we find $x_1=\frac{5}{48},$ $x_1^S=\frac{5}{8},$ and $x_1^S({\rm sp})=-\frac{1}{24}.$
From (\ref{eq:key}) we immediately have expressions for the exponents we discussed above. In particular,  $$\gamma=\nu[2-2x_1]=43/32,$$ $$\gamma_1=\nu[2-x_1-x_1^S]=\frac{61}{64},$$ and 
\begin{equation}\label{g11}
\gamma_{11}=\nu[1-2x_1^S]=-\frac{3}{16}.
\end{equation}
Similarly, at the special transition, we immediately obtain $$\gamma_1({\rm sp})=\nu[2-x_1-x_1^S({\rm sp})]=\frac{93}{64},$$ and 
\begin{equation}\label{g11sp}
\gamma_{11}({\rm sp})=\nu[1-2x_1^S({\rm sp})]=\frac{13}{16}.
\end{equation}
 These results imply the Barber scaling relation above, and furthermore its extension to the exponents at the special transition.

Notice that formula \key  cannot, {at first sight}, predict the exponent for bridges, as there {seems to be} no way to specify the constraint that the end point has maximal displacement in the direction normal to the surface hyper-plane. {However, by considering networks confined between two parallel hyper-planes, a simple reinterpretation and extension  of \key will allow us}  to address this problem, and we do so in the next section.

\section{Extension to bridge-like configurations.}
Before deriving a result for the critical exponent of bridges, we will rederive a known result, that of the critical exponent characterising self-avoiding polygons. The number of $2n$-step polygons on a hyper-cubic lattice is expected to behave asymptotically as $$p_{2n} \sim F \cdot \mu^{2n} \cdot n^{\alpha-3},$$ so the generating function has exponent $2 - \alpha.$
SAPs anchored at a surface can be considered in two different ways, as shown in Fig \ref{fig:poly}. Firstly, as a single loop, anchored at the origin (Fig \ref{fig:poly}(a)). In that case we have from \key $\V_S=1, \, \N=1, \, n_2=1.$ It is also known \cite[Section 6.5.1]{D89} (see also \cite{DDE83}) that in any dimension $x_2^S=d.$ Thus we find $$\gamma_p = \nu[-x_2^S]=-\nu d.$$ This is just the well-known hyper-scaling relation $\nu d=2 - \alpha.$

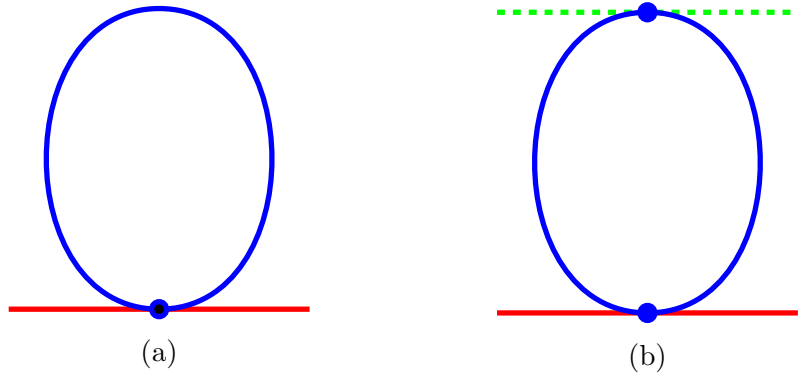
\begin{figure}
\centering
\begin{subfigure}{0.49\textwidth}
\centering
\begin{tikzpicture}
\tikzset{vert/.style={line width=2pt, circle, fill=black, draw=blue, inner sep=2pt}}
\draw [line width=2pt, red] (0,0) -- (4,0);
\draw [line width=2pt, blue] (2,0) node [vert] {} .. controls (0,0) and (0,4) .. (2,4) .. controls (4,4) and (4,0) .. (2,0);
\end{tikzpicture}
\caption{}
\end{subfigure}
\begin{subfigure}{0.32\textwidth}
\centering
\begin{tikzpicture}
\tikzset{vert/.style={line width=2pt, circle, fill=blue, draw=blue, inner sep=2pt}}
\draw [line width=2pt, red] (6,0) -- (10,0);
\draw [dashed, line width=2pt, green] (6,4) -- (10,4);
\draw [line width=2pt, blue] (8,0) node [vert] {} .. controls (6,0) and (6,4) .. (8,4) node [vert] {} .. controls (10,4) and (10,0) .. (8,0);
\end{tikzpicture}
\caption{}
\end{subfigure}
\caption{A polygon as a loop anchored in the surface (a), and as a watermelon with two surface vertices, with the surface (dotted line) at the ring's top-most point being virtual (b).}
\label{fig:poly}
\end{figure}

Another way of viewing the polygon is as a two-armed watermelon, with the top-most vertex being unconstrained (Fig \ref{fig:poly}(b)), provided it lies in a parallel surface at maximal spacing between the two surfaces.  So from \key one has $\N=2,$ $\V=1,$ $\V_S=1,$ $n_S=2.$
This gives $$\gamma_p' = \nu[d-2 \cdot x_2^S]-1=-\nu d -1.$$ 
Now the top-most vertex can be any of the $n$ vertices of the polygon {({\it i.e.}, seen here as a \emph{polydisperse} 2-leg watermelon)}, so taking this into account the correct exponent for polygons is $\gamma_p' +1=\gamma_p.$ Note that $\gamma_p=-\nu d,$ obtained here for a {\em surface anchored} SAP, is the same hyperscaling relation $\nu d = 2-\alpha$ as expected for a bulk SAP. The reason is exactly the same as before: the anchoring surface can also be seen as a virtual one, this time marking the bottom-most point of a bulk SAP, which does not change the configurational exponent of the SAP.

\begin{figure}[htb]
\centering
\includegraphics[scale =0.90] {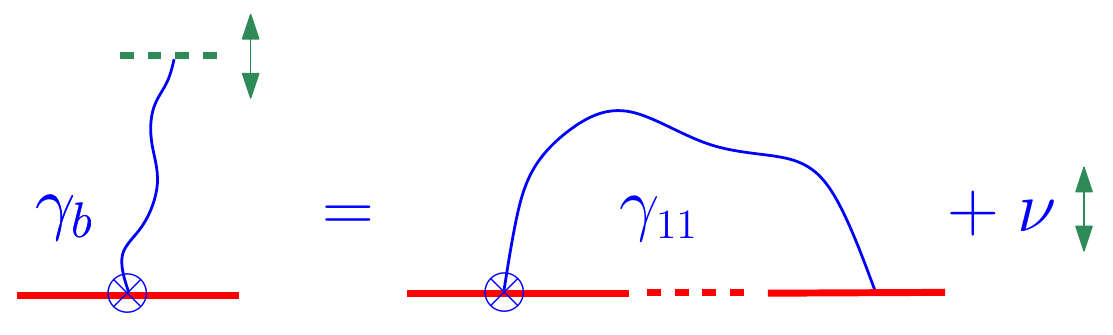}
 \caption{Identity \eqref{eq:gbnug11} viewed as resulting from a ``vertex algebra'':   $\otimes$ represents a {1-vertex} fixed on the surface, while a line touching a (continuous or dotted) segment represents a (real or virtual) movable surface  The double-headed arrow represents free motion along the vertical direction,  which yields a contribution $\nu$ to the bridge configurational exponent.}
 \label{fig:bridge3}
\end{figure}
Turning now to bridges, a bridge can be considered as a SAW rooted at the surface, but with end-point free to move in the bulk, provided it lies in a parallel surface at maximal spacing between the two surfaces, just as the second vertex of the polygon previously considered. In this way we obtain from (\ref{eq:key}) that bridges can be described by networks with ${\mathcal V}=1,$ ${\mathcal V}_S=1,$ $\N=1,$ $n_1=0,$ and $n_1^S=2.$ This gives the exponent for bridges as 
\begin{equation}\label{eq:gbnug11}\gamma_b=\nu[2-2x_1^S]=\gamma_{11}+\nu.
\end{equation}
This identity is easily explained in graphical terms (Fig. \ref{fig:bridge3}). In two dimensions it gives $\gamma_b = \frac{9}{16},$ as expected.

This is a new scaling relation, and it can perhaps be more appealingly written as 
$$\gamma_1 =\frac{1}{2}(\gamma + \gamma_b).$$ It is obvious that $\gamma_1$ is bounded above by $\gamma,$ and below by $\gamma_b.$ What is perhaps surprising is that it is precisely the average of these two exponents. However, this identity can also be rewritten as 
\begin{equation}\label{eq:gg1gb}
\gamma-\gamma_1 =\gamma_1 -\gamma_b,
\end{equation}  
which then allows for a simple graphical explanation. 
\begin{figure}[h!]
\centering
\includegraphics[scale =0.98] {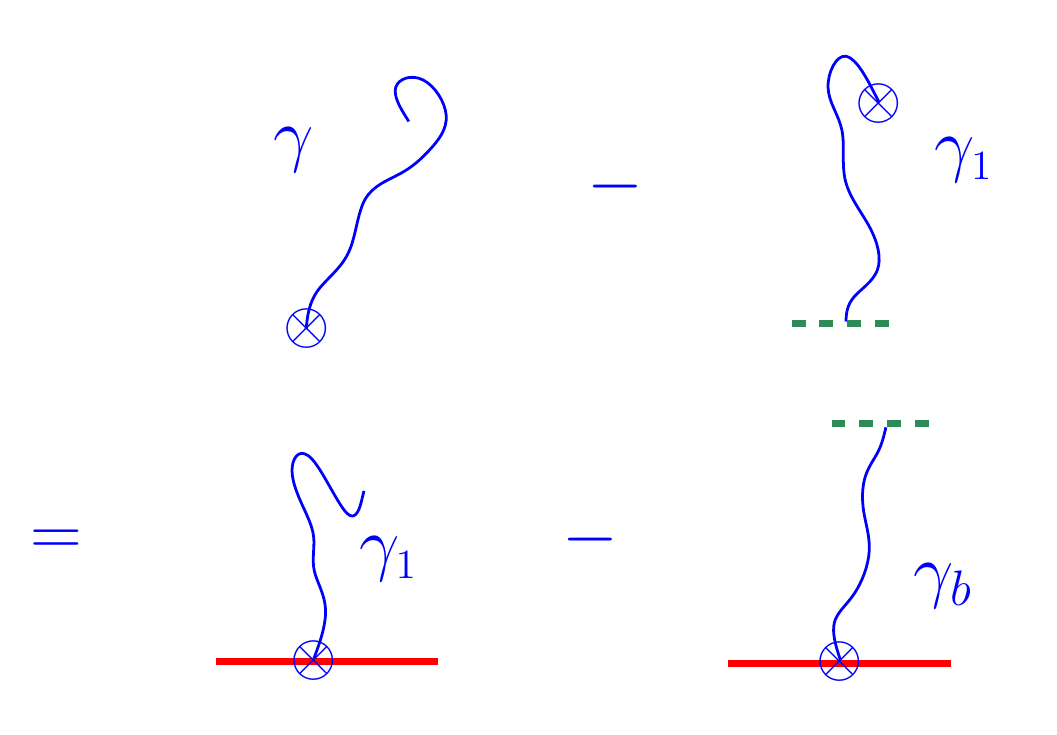}
 \caption{{Identity \eqref{eq:gg1gb} viewed as resulting from the above mentioned vertex algebra:   $\otimes$ represents a {1-vertex} fixed in space, {a free extremity represents a 1-vertex floating in the bulk,} while a line anchored at a (continuous or dotted) segment represents a (real or virtual) surface vertex.}}
 \label{fig:bridge2}
\end{figure}

In Fig. \ref{fig:bridge2}, the cancellation of identical vertices,  performed separately on the left and right members of the equation, leaves the same set of remaining vertices on both sides of the equality, namely one bulk vertex and one surface vertex. Eq. \eqref{eq:key} then yields $\gamma-\gamma_1=\nu(x_1^S-x_1)=\gamma_1-\gamma_b$. 

\subsection{The special transition}
For the special transition, the origin vertex is at critical fugacity, so one of the two surface exponents $x_1^S$ has to be replaced by its value $x_1^S({\rm sp})$ at the special transition. In this way we find
$$\gamma_b({\rm sp})=\nu[2-x_1^S - x_1^S({\rm sp})].$$ 
Recalling  Eqs. \eqref{g11} and \eqref{g11sp}, 
we readily obtain the special transition scaling relation, 
\begin{equation}\label{eq:gbspg11}
\gamma_b({\rm sp})=\frac{1}{2}\left[\gamma_{11}({\rm sp})+\gamma_{11}\right] + \nu.
\end{equation}
\begin{figure}[h!]
\centering
\includegraphics[scale =0.80] {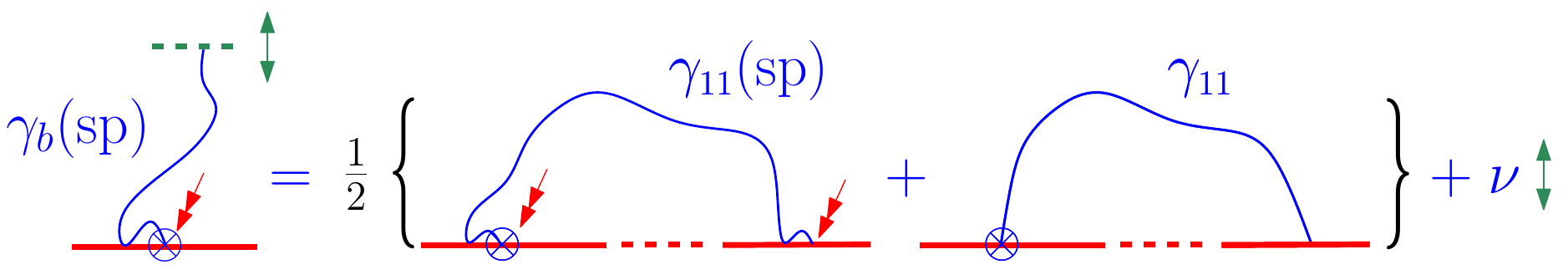}
 \caption{Identity \eqref{eq:gbspg11} viewed as resulting from the extended vertex algebra:   All elements are as in Figs. \ref{fig:bridge3} and \ref{fig:bridge2}, with the novel surjective arrow representing a surface vertex at the special transition point. The numbers of occurrences of various types of surface vertices (fixed or free, at the ordinary or special transitions) are the same on both sides of the equation, implying equality of exponents.}
 \label{fig:gbspg11}
\end{figure}
Again, this relation is easily understood in graphical terms (Fig. \ref{fig:gbspg11}). 

In two dimensions we have from (\ref{eq:xsurf}) that $x_1^S({\rm sp}) = -1/24,$ so this gives for bridges $\gamma_b({\rm sp})=17/16.$ Similarly, for polygons at the critical transition, one has 
$\gamma_p({\rm sp}) = \nu[-x_2^S({\rm sp})].$  In two dimensions we have from (\ref{eq:xsurf}) that $x_2^S({\rm sp}) = 1/3,$  so in two dimensions, for SAPs at the special transition point, $\gamma_p({\rm sp})=-1/4.$

\subsection{Three-dimensional TAWs and bridges}
In three dimensions the most precise results we have are from Monte Carlo analysis. Clisby \cite {C10, C17} has given the estimates $\nu = 0.587597 \pm 0.000007,$ and $\gamma = 1.156957 \pm 0.000009.$ For $\gamma_1$ there are a few estimates in the literature based on rather short series. Some 30 years ago, Guttmann and Torrie \cite{GT84} estimated $\gamma_1 = 0.676 \pm 0.009,$ while one of the few Monte Carlo estimates, already 20 years old, is by Hegger and Grassberger \cite{HG94} who estimated $\gamma_1= 0.679 \pm 0.002.$

More recently Clisby, Conway and Guttmann \cite{CCG16} studied both TAWs and bridges on the simple-cubic lattice, using both series and Monte Carlo methods. The series results, based on the analysis of a 26-term series for TAWs and a 28-term series for bridges, and utilising Clisby's precise estimate of the critical point \cite{C13}, were $\gamma_1 = 0.676 \pm 0.002,$ and $\gamma_b = 0.199 \pm 0.002.$ The Monte Carlo  estimates were much more precise, being $\gamma_1 = 0.677667 \pm 0.000017,$ and $\gamma_b = 0.198352 \pm 0.000027.$
 In four dimensions all exponents are known to take their mean-field value (with logarithmic corrections), and the scaling relations are trivially satisfied.

From the epsilon expansions of  Diehl and Dietrich \cite{DD80} and Reeve and Guttmann \cite{RG81} for the $O(n)$-model (see also \cite{Dieh86}), it follows from the scaling relation \eqref{eq:gbnug11} that $$\gamma_b = \frac{(n+2)}{2(n+8)}\varepsilon + \frac{(n+2)(n^2+27n+92)}{4(n+8)^3}\varepsilon^2 + O(\varepsilon^3),$$ where as before $\varepsilon=d-4.$ For $n=0$ this evaluates   to $\gamma_b = 0.6093\ldots, 0.2148\ldots, 0$ for $d=2,3,4$ respectively. These results are quite close to the numerical values given above.

\subsection{Bridge exponent at the special transition in three dimensions}
To calculate the bridge exponent at the special transition we require the epsilon expansion for $\gamma_{11}$ at the special transition. This was obtained by Reeve \cite{R81} and Diehl and Dietrich \cite{DD81} (see also \cite{Dieh86}) for the $O(n)$-model} as $$\gamma_{11}({\rm sp}) = \frac{1}{2}  + \frac{3(n+2)}{4(n+8)}\varepsilon +\frac{(n+2)(3n^2+53n+52)}{8(n+8)^3}\varepsilon^2+ O(\varepsilon^3),$$ so from our scaling relation \eqref{eq:gbspg11} we find the epsilon expansion for the bridge exponent at the special transition to be
$$\gamma_{b}({\rm sp}) = \frac{1}{2}  + \frac{3(n+2)}{4(n+8)}\varepsilon +\frac{(n+2)(3n^2+65n+148)}{8(n+8)^3}\varepsilon^2+ O(\varepsilon^3),$$
which for $n=0$ evaluates to $389/512 \approx 0.760$ at $\varepsilon=1.$

Monte Carlo calculations of $\gamma_1({\rm sp})$ combined with the scaling relation (\ref{barb}), also valid at the special transition point,  allows us to estimate $\gamma_{11}({\rm sp}).$ In \cite{G05} Grassberger gives the estimate $\gamma_1({\rm sp})= 1.226 \pm 0.002,$ while more recently Klushin et al. \cite{K13} give $\gamma_1({\rm sp})= 1.224 \pm 0.003.$ Older results from massive field theory by Diehl and Shpot \cite{DieS94,DieS98},  give $\gamma_1({\rm sp})\approx 1.207.$ We will take as our estimate the mean of the two recent Monte Carlo calculations, so that $\gamma_1({\rm sp}) \approx 1.225,$ and from eqn. (\ref{barb}) this gives $\gamma_{11}({\rm sp}) \approx 0.705,$ and so $\gamma_b({\rm sp})\approx 0.746.$ This is in surprisingly good agreement with the epsilon expansion result given in the previous paragraph.

For completion, we remark that  conformal bootstrap methods \cite{GLMR} have recently been used to estimate polymer surface critical exponents, 
confirming the results of Refs. \cite{DieS94,DieS98} for the ordinary transition, but they have not generated reliable results for exponents at the special transition.

\section{Scaling law for worm configurations}
Another subset of SAWs that is of interest are worms, as defined above. The exponent $\gamma_w$ can be predicted from Eqn. (\ref{eq:key}) if we take the two end-points to be surface vertices, so that $\V_s=2,$ and $\V=0.$ However the vertices are otherwise unconstrained, so that $n_1=2$ and $n_1^S=0.$ This gives $$ \gamma_w=\nu(1-2x_1)=\gamma-\nu.$$

Alternatively, we can give a simple geometric  argument for the scaling relation $\gamma_w = \gamma - \nu.$  The exponent for SAWs is $\gamma.$ The end-point of a SAW is assumed to be radially symmetric, modulo lattice effects. The average length of an $n$-step SAW is proportinal to $n^\nu,$ and so the proportion of SAWs ending on any particular radial line scales as $n^{-\nu}.$ Taking the radial line as the $x$ axis gives the exponent for worms as $\gamma - \nu.$ Thus for two-dimensional worms, we expect an exponent of $$\gamma_w = \gamma - \nu =43/32-3/4=19/32.$$ Iwan Jensen (private communication) has calculated the end-point distribution of square-lattice SAWs up to length 59. We can extract the coefficients of the worm generating function to the same order from this data, and series analysis we have performed confirms this prediction to five-digit precision. We have not carried out any enumerations for three-dimensional lattice worms, but the scaling argument given implies $\gamma_w(3d)=0.56936 \pm 0.000016.$

\section{Conclusion}
 We have given several examples showing how the theory of the critical behaviour of $d$-dimensional polymer networks {\cite{BD86,DS86,D89} can be extended to the situation of bridges where the chains lie between two parallel hyper-planes. In this way we have derived a new scaling relation for self-avoiding bridges that relates the critical exponents of bridges and terminally-attached self-avoiding arches, and the length exponent $\nu.$ We also give supportive results based on series and Monte Carlo enumeration data. Unlike many scaling laws, this requires modification when describing the special transition, and the appropriately modified scaling law is also derived.
 
 We have also derived a scaling relation for a subset of SAWs called worms. This has been verified numerically in the two-dimensional case by series analysis. It is also possible to extend the theory  more generally to polymer networks between parallel hyperplanes, {as well as to the case of the tricritical polymer $\Theta$-point,} and this will be the subject of a future article.
\section*{Acknowledgements}
We wish to acknowledge the hospitality of the Erwin Schr\"odinger International Institute for Mathematical Physics where this work was initiated, during the programme on {\em Combinatorics, Geometry and Physics} in June, 2014.  AJG wishes to thank the Australian Research Council for supporting this work through grant DP120100931, and more recently ACEMS, the ARC Centre of Excellence for Mathematical and Statistical Frontiers. We also wish to warmly thank Hans Werner Diehl for pointing out a number of references relevant to surface transitions, and Emmanuel Guitter for his kind help with the figures.

\end{document}